%% file: main.tex
\documentclass[lettersize,journal]{IEEEtran}
\usepackage{amsmath,amsfonts}
\usepackage{algorithmic}
\usepackage{algorithm}
\usepackage{array}
\usepackage[caption=false,font=footnotesize]{subfig}
\usepackage{textcomp}
\usepackage{stfloats}
\usepackage{url}
\usepackage{verbatim}
\usepackage{graphicx}
\usepackage{cite}
\usepackage{booktabs}
\usepackage{multirow}
\usepackage{adjustbox}
\usepackage{pifont}
\usepackage[hidelinks]{hyperref}
\newcommand{\cmark}{\ding{51}}
\newcommand{\xmark}{\ding{55}}
\graphicspath{{figures/}}
\begin{document}

\title{Cross-Lingual F5-TTS 2: A Simplified Framework for Language-Agnostic Voice Cloning}

\author{Qingyu Liu*, Rixi Xu*, Yushen Chen, Zhikang Niu, Haitao Li,
        Pengcheng Zhu, Bowen Zhang, Jian Zhao, Yunting Yang, Qinyuan Cheng, Xipeng Qiu, Berrak Sisman, Kai Yu, and Xie Chen%
\thanks{Qingyu Liu and Berrak Sisman are with the Center for Language and Speech Processing, Johns Hopkins University, Baltimore, MD, USA.}%
\thanks{Qingyu Liu, Rixi Xu, Yushen Chen, Zhikang Niu, Kai Yu, and Xie Chen are with the MoE Key Lab of Artificial Intelligence, X-LANCE Lab, Shanghai Jiao Tong University, Shanghai, China. (e-mail: qliu91@jh.edu, chenxie95@sjtu.edu.cn)}%
\thanks{Yushen Chen, Zhikang Niu, Haitao Li, Qinyuan Cheng, Xipeng Qiu, and Xie Chen are with Shanghai Innovation Institute, Shanghai, China.}%
\thanks{Pengcheng Zhu, Bowen Zhang, Jian Zhao, and Yunting Yang are with Geely, China.}%
\thanks{Haitao Li is with Zhejiang University, Zhejiang, China.}%
\thanks{Qinyuan Cheng and Xipeng Qiu are with Fudan University, Shanghai, China.}%
\thanks{Qingyu Liu and Rixi Xu contributed equally to this work.}%
\thanks{Xie Chen is the corresponding author.}%
}

\markboth{IEEE/ACM Transactions on Audio, Speech, and Language Processing}%
{Liu \MakeLowercase{\textit{et al.}}: Cross-Lingual F5-TTS 2}

\maketitle

\input{sections/0_abstract}


\input{sections/1_introduction}
\input{sections/2_related_work}
\input{sections/3_method}
\input{sections/4_experiments}
\input{sections/5_results}
\input{sections/6_conclusion}


\bibliographystyle{IEEEtran}
\bibliography{references}

\end{document}

%% file: sections/0_abstract.tex
\begin{abstract}
Zero-shot text-to-speech (TTS) can clone a speaker's voice from a short audio prompt, yet most TTS systems still require the audio prompt transcript during inference. This dependency prevents cross-lingual voice cloning when the audio prompt transcript is unavailable, particularly for unseen languages. Cross-Lingual F5-TTS removes this dependency and enables transcript-free cross-lingual voice cloning, but it prepares its training data with forced alignment. Forced alignment is sensitive to boundary errors, and its cost grows as more languages are covered. Its speaking rate predictor is also unreliable at estimating duration when the audio prompt begins or ends with silence.
In this paper, we present Cross-Lingual F5-TTS 2, a simplified framework for transcript-free cross-lingual voice cloning without forced alignment. Instead of using forced alignment to segment real utterances, we build same-speaker prompt and target pairs using a pretrained F5-TTS model and fine-tune the same model on these constructed pairs. This simplifies data preparation and preserves the acoustic modeling capability of the pretrained model, enabling adaptation with only a short fine-tuning stage. We further make the syllable-level speaking rate predictor robust to leading and trailing silence through silence-aware augmentation. Experiments show that Cross-Lingual F5-TTS 2 reaches higher speaker similarity than F5-TTS and Cross-Lingual F5-TTS while maintaining intelligibility. All related resources are publicly available.\footnote{\url{https://qingyuliu0521.github.io/Cross-Lingual_F5-TTS_2_demo/}}

\end{abstract}

%% file: sections/1_introduction.tex
\section{Introduction}
\IEEEPARstart{Z}{ero-shot} text-to-speech (TTS), also known as voice cloning, aims to generate speech that closely resembles a target speaker's voice from only a brief audio prompt. Recent systems fall into three families. Autoregressive (AR) codec language models such as VALL-E~\cite{wang2023valle} treat speech generation as next-token prediction and pioneered in-context voice cloning. Hybrid systems such as the CosyVoice~\cite{du2024cosyvoice, du2024cosyvoice2, du2025cosyvoice3} and IndexTTS~\cite{zhou2025indextts2, li2026indextts25} series combine AR semantic modeling with non-autoregressive (NAR) acoustic refinement in a coarse-to-fine framework. Purely NAR flow-matching models such as Voicebox~\cite{le2023voicebox}, E2-TTS~\cite{eskimez2024e2tts}, and F5-TTS~\cite{chen2025f5tts} instead generate all acoustic frames in parallel, which avoids the sequential bottleneck of AR decoding and yields faster synthesis with a streamlined architecture that needs no cascaded discrete-token pipeline. Across these families, however, most systems still condition on the transcript of the audio prompt during inference.

This transcript dependency is particularly difficult to avoid in cross-lingual voice cloning. When the reference speaker and the target text are in different languages, the prompt transcript is often unreliable or simply unavailable, especially when a user provides spontaneous speech in a low-resource or unfamiliar language. A transcript-free method is therefore most valuable in this setting, provided it still keeps the full in-context prompt.

Two main lines of work have removed this dependency. One line conditions on a global speaker embedding instead of an in-context prompt, as in YourTTS~\cite{casanova2022yourtts}; Qwen3-TTS~\cite{hu2026qwen3tts} likewise offers a speaker-embedding mode for transcript-free cloning. Such embeddings sidestep the transcript but keep only a coarse copy of the speaker and lose the local timbre and prosody that in-context prompting carries over. The other line keeps in-context prompting and builds same-speaker prompt-target pairs during data preparation, as in MOSS-TTS~\cite{gong2026mosstts} and the IndexTTS series~\cite{zhou2025indextts2, li2026indextts25}, which removes the transcript at inference at the cost of a heavy preprocessing pipeline.

Removing the transcript is more involved for F5-TTS. Instead of explicit phoneme alignment or a separate duration predictor, F5-TTS pads the character sequence with filler tokens and estimates the target duration from the length ratio between the audio prompt transcript and the target text~\cite{chen2025f5tts}. The transcript therefore serves two roles: it forms the prompt portion of the text the model conditions on, and its length sets the reference for duration estimation. Without it, both the text condition and the duration estimate break down.

Cross-Lingual F5-TTS~\cite{liu2026crosslingualf5tts} addresses both roles of the transcript. For the text condition, it uses forced alignment to split each training utterance into a prompt and a target segment and discards the prompt transcript, so that inference no longer requires it; VoXtream2~\cite{torgashov2026voxtream2} likewise relies on forced alignment for training-data segmentation. For duration, because NAR models must fix the target length before generation and can no longer derive it from a text length ratio, it trains a syllable-level speaking rate predictor (SRP) over the audio prompt. However, both components leave residual problems. Forced alignment is sensitive to boundary errors and grows more expensive as language coverage widens. The speaking rate predictor becomes unreliable when the audio prompt carries leading or trailing silence: the silence lowers the estimated speaking rate, so the predicted duration is inflated and the generated speech is stretched, which hurts naturalness and intelligibility.

In this paper, we propose Cross-Lingual F5-TTS 2, a simplified supervised fine-tuning (SFT) framework built on F5-TTS that removes the audio prompt transcript at inference without forced alignment. For each training sample, a pretrained F5-TTS model synthesizes an audio prompt in the same speaker's voice, which we concatenate with the real speech as the target. We fine-tune on these pairs without the audio prompt transcript, matching transcript-free inference.

This paper makes the following contributions:
\begin{itemize}
  \item \textbf{Simplified Supervised Fine-Tuning Framework:} We propose an SFT framework built on F5-TTS that pairs each real utterance with a same-speaker prompt synthesized by the pretrained model, removing the audio prompt transcript at inference and simplifying data preparation.
  \item \textbf{Prompt-Token Text Conditioning:} We introduce ratio-based learnable prompt tokens with an end-of-sequence (EOS) marker that preserve the pretrained text-audio layout and accelerate WER convergence during fine-tuning.
  \item \textbf{Silence-Robust Speaking Rate Predictor:} We make transcript-free duration estimation robust to leading or trailing silence in the audio prompt through silence-aware augmentation.
\end{itemize}

The remainder of this paper is organized as follows. Section II reviews related work. Section III presents the proposed method. Section IV describes the experimental setup, followed by results and analysis in Section V. Section VI concludes the paper.

%% file: sections/2_related_work.tex
\section{Related Work}
\label{sec:related_work}
\subsection{Transcript-Free Voice Cloning}

Most zero-shot TTS systems require the transcript of the audio prompt at inference to condition generation on the reference speaker.

A widely used approach conditions generation on a global speaker embedding. YourTTS~\cite{casanova2022yourtts} represents the reference with a single speaker vector at both training and inference, and Qwen3-TTS~\cite{hu2026qwen3tts}, though trained with in-context learning, additionally offers a speaker-embedding mode that needs no prompt transcript. A speaker embedding removes the transcript directly, but it retains only a coarse speaker identity and loses the local timbre and prosody that an in-context prompt preserves.

Another approach keeps in-context prompting and obtains same-speaker prompt-target pairs during data preparation. MOSS-TTS~\cite{gong2026mosstts} and the IndexTTS series~\cite{zhou2025indextts2, li2026indextts25} build such pairs either by mining open-domain recordings with speaker diarization and single-speaker segment merging, or by drawing on speaker-labeled corpora such as Emilia~\cite{he2024emilia}. This preserves in-context fidelity, but removes the audio prompt transcript only through a heavy data-preparation pipeline.

More recent systems rely on a forced aligner when preparing the training data. Cross-Lingual F5-TTS~\cite{liu2026crosslingualf5tts} uses MMS forced alignment~\cite{pratap2024mms} to locate word boundaries, splits each utterance into a prompt and a target segment, and discards the prompt transcript. VoXtream2~\cite{torgashov2026voxtream2} depends on the Clap-IPA forced aligner to align phonemes to audio frames. Forced alignment is error-prone and wastes data: VoXtream2 discards about 35\% of its training corpus after removing failed alignments and invalid transcripts.

Removing the audio prompt transcript thus comes at a cost along every route, either in speaker fidelity or in the effort and data loss of training-data preparation. This leaves open the need for a transcript-free method that preserves full in-context fidelity without a costly data preparation pipeline.

\subsection{Synthetic Data for TTS Training}

Synthetic speech generated by existing models is increasingly used to train TTS systems. One direction focuses on the synthetic-data pipeline itself: a study of training with purely synthetic data~\cite{synthtts2025} shows that diverse, clean synthetic speech can match or even surpass real recordings, and SpeechWeave~\cite{speechweave2025} generates multilingual, speaker-standardized text-speech pairs to improve data diversity and normalization. These works establish synthetic speech as a viable general-purpose training resource.

Another direction uses synthetic data to compensate for scarce real data, particularly for low-resource languages. LLM-to-Speech~\cite{niletts2026} builds an Egyptian Arabic corpus by generating dialectal text with an LLM and converting it to speech, then fine-tunes a multilingual TTS model on it, while a recent study on low-resource spoken language modeling~\cite{slmgap2026} uses synthetic supervision to enable zero-shot voice cloning for Lao. 


\subsection{Duration Modeling for NAR Zero-Shot TTS Systems}

Non-autoregressive zero-shot TTS systems must determine the target acoustic length before acoustic generation. Earlier zero-shot NAR systems model an explicit phoneme-level duration and obtain the per-phoneme duration targets from an alignment tool. Voicebox~\cite{le2023voicebox} decouples acoustic and duration modeling, training a separate duration model and relying on a forced aligner for the duration targets. NaturalSpeech 3~\cite{ju2024naturalspeech3} instead models duration with a diffusion module and a length regulator, using an internal alignment tool to obtain the phoneme-level durations. Both systems couple the duration stage to an external aligner during data preparation.

Recent NAR TTS systems avoid this dependency. E2-TTS~\cite{eskimez2024e2tts} and F5-TTS~\cite{chen2025f5tts} pad the character sequence with filler tokens to the mel length, so they need neither frame-level text-speech alignment nor a phoneme-level duration predictor. At inference, however, the total target duration still has to be set. F5-TTS infers it from the length ratio between the reference and target text, and M3-TTS~\cite{m3tts2025}, though it replaces filler padding with cross-modal attention, likewise sets the output length from a reference speech-to-text ratio. The ratio is simple and needs no extra predictor, but it breaks down across languages, where a text-length ratio need not match the speech-duration ratio. Cross-Lingual F5-TTS~\cite{liu2026crosslingualf5tts} removes this limitation by estimating a speaking rate from the audio prompt rather than from a text-length ratio, so the estimate transfers across languages.

Whether from a text-length ratio or a speaking rate predictor, these methods derive the target duration from the audio prompt, assuming the prompt is speech throughout. However, audio prompts supplied by users often carry leading or trailing silence, which lowers the apparent speaking rate, inflates the estimated duration, and stretches the synthesized speech.

%% file: sections/3_method.tex
\begin{figure*}[!t]
  \centering
  \includegraphics[width=0.8\textwidth]{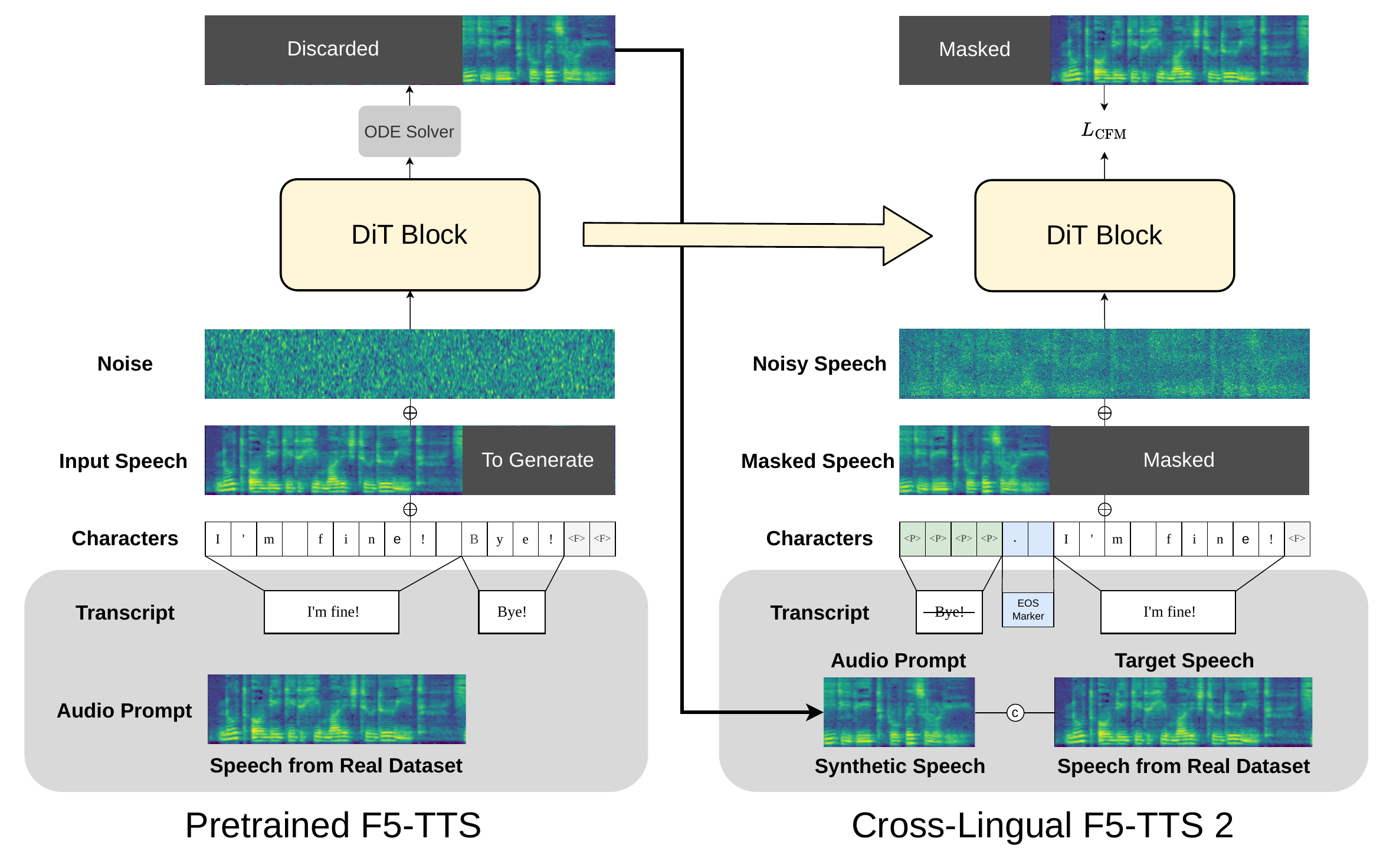}
  \caption{Overview of the proposed SFT pipeline. 
  (Left) \textbf{Synthetic Data Generation}: a pretrained F5-TTS model is used to synthesize speaker-consistent audio pairs. 
  (Right) \textbf{SFT with Synthetic Prompt}: Cross-Lingual F5-TTS 2 is initialized from pretrained F5-TTS and fine-tuned on the constructed real-synthetic pairs, where the audio prompt transcript is replaced by learnable prompt tokens followed by an EOS marker, forcing the model to condition on acoustic prompts without relying on their transcripts.}
  \label{fig:synthetic_prompt_sft}
\end{figure*}

\section{Proposed Method}
\label{sec:method}

We propose Cross-Lingual F5-TTS 2, a transcript-free fine-tuning framework built on pretrained F5-TTS.
Instead of using MMS forced alignment to remove audio prompt transcripts, the proposed framework uses synthetic audio prompts to form real-synthetic pairs, and replaces the missing prompt transcript with learnable prompt tokens followed by an end-of-sequence (EOS) marker. The overall SFT pipeline is shown in Fig.~\ref{fig:synthetic_prompt_sft}.
This section first reviews the preliminaries on flow matching, F5-TTS, and Cross-Lingual F5-TTS, then describes the SFT procedure, and finally introduces the silence-robust speaking rate predictor used for duration estimation.

\subsection{Preliminaries}

\textbf{1) Flow Matching:}

The Flow Matching (FM) objective is to learn a time-dependent vector field $v_t$ that matches a probability path $p_t$ from a simple prior distribution $p_0$ to the data distribution $q$~\cite{lipman2023flowmatching}.
The FM loss regresses $v_t$ against the target field $u_t$:
\begin{equation}
  \mathcal{L}_{\mathrm{FM}}(\theta)
  = \mathbb{E}_{t,\,p_t(x)} \left\| v_t(x;\theta) - u_t(x) \right\|^2.
\end{equation}
Since $p_t$ and $u_t$ are intractable, a conditional probability path $p_t(x|x_1)$ is used in practice, yielding the Conditional Flow Matching (CFM) loss with identical gradients:
\begin{equation}
  \mathcal{L}_{\mathrm{CFM}}(\theta)
  = \mathbb{E}_{t,\,q(x_1),\,p(x_0)}
    \left\| v_t\!\left(\psi_t(x_0);\theta\right) - \frac{\mathrm{d}}{\mathrm{d}t}\psi_t(x_0) \right\|^2.
\end{equation}
Further leveraging the Optimal Transport (OT) form $\psi_t(x_0) = (1-t)x_0 + tx_1$, the OT-CFM loss simplifies to
\begin{equation}
\resizebox{0.9\linewidth}{!}{$\displaystyle
  \mathcal{L}_{\mathrm{OT\text{-}CFM}}(\theta)
  = \mathbb{E}_{t,\,q(x_1),\,p(x_0)}
    \left\| v_t\!\left((1-t)x_0+tx_1;\theta\right) - (x_1-x_0) \right\|^2,
$}
\end{equation}
where the target velocity field reduces to the constant vector $(x_1-x_0)$.

\textbf{2) F5-TTS:}

F5-TTS~\cite{chen2025f5tts} adopts a Diffusion Transformer (DiT) backbone and is trained on the text-guided speech infilling task.
Given surrounding speech $(1-m)\odot x_1$, a noisy mixture $(1-t)x_0+tx_1$, and a filler-token-padded character sequence, the model predicts the masked target region $m\odot x_1$.
Filler tokens pad the character sequence to match the mel-frame length, eliminating the need for explicit phoneme alignment or a separate duration predictor.

At inference, the target mel spectrogram is obtained by integrating the predicted vector field from an initial noise sample $x_0$ using an ODE solver:
\begin{equation}
  x_1 = x_0 + \int_0^1 v_t\!\left(\phi_t(x_0),\,c;\,\theta\right)\mathrm{d}t.
\end{equation}

Target duration is estimated from the length ratio of the audio prompt transcript to the target text, which approximates the ratio of prompt to target mel frames.
The audio prompt transcript thus plays a dual role: it contributes to the text-side condition observed by the model, and provides the length reference for duration estimation.

\textbf{3) Cross-Lingual F5-TTS:}

Cross-Lingual F5-TTS~\cite{liu2026crosslingualf5tts} extends F5-TTS to remove the audio prompt transcript dependency.
During training, MMS forced alignment~\cite{pratap2024mms} extracts word boundaries from each utterance; a randomly selected boundary partitions it into a prompt and a target segment, after which the prompt transcript is discarded and only the target transcript is retained as the text condition.
At inference, a speaking rate predictor estimates the target duration directly from the audio prompt, replacing the transcript-dependent text length ratio.

The speaking rate predictor is a transformer-based model that takes mel-spectrograms as input and predicts discrete speaking rate categories using Gaussian Cross-Entropy (GCE) loss, which incorporates the ordinal structure of the rate intervals via Gaussian soft labels:
\begin{equation}
  L_{\text{GCE}}
  = -\frac{1}{N_s}\sum_{i=1}^{N_s}\sum_{c=1}^{C}
  y^{\text{soft}}_{c} \log(\hat{y}_c),
  \label{eq:related_gce}
\end{equation}
where the soft label for category $c$ is computed by a Gaussian kernel centered at the ground truth category $c_{\text{gt}}$:
\begin{equation}
  y_c^{\text{soft}}
  = e^{\frac{-(c-c_{\text{gt}})^2}{2\sigma^2}}.
  \label{eq:related_ysoft}
\end{equation}
The standard deviation $\sigma$ controls the smoothness of the soft label distribution, assigning larger weights to speaking-rate categories closer to the ground truth.
Cross-Lingual F5-TTS trains predictors at phoneme, syllable, and word granularities, and finds that the syllable-level predictor generalizes best in cross-lingual settings, where the audio prompt comes from a language unseen during predictor training.


\subsection{Synthetic-Prompt Supervised Fine-Tuning}

\textbf{1) Real-Synthetic Pair Construction:}

Rather than relying on forced alignment to partition real speech, we construct prompt-target pairs using a pretrained F5-TTS model.
For each real training sample, the sample itself serves as the audio prompt, and a target text is drawn from a text pool in the corresponding language.
The pretrained model synthesizes a new utterance in the same speaker's voice, yielding a synthetic audio prompt $x_{\text{prompt}}^{\text{syn}}$.
This synthetic prompt is paired with the original real speech as the target $x_{\text{target}}$, forming a complete training pair without requiring an external forced aligner.

\textbf{2) SFT with Synthetic Prompt:}

We retain the speech infilling formulation of F5-TTS but remove the transcript associated with the synthetic audio prompt.
The acoustic input $x_1$ is constructed by concatenating the synthetic audio prompt and the real target speech along the temporal axis:
\begin{equation}
  x_1 = \bigl[x_{\text{prompt}}^{\text{syn}};\, x_{\text{target}}\bigr].
\end{equation}
Let $\tau_1$ and $\tau_2$ denote the frame lengths of the prompt and target respectively, and $\tau = \tau_1 + \tau_2$ the total length. A binary temporal mask $m$ is then defined as
\begin{equation}
  m[:,\, j] =
  \begin{cases}
    0, & j \in [0,\, \tau_1), \\
    1, & j \in [\tau_1,\, \tau),
  \end{cases}
\end{equation}
so that $(1-m)\odot x_1$ exposes the synthetic audio prompt as the acoustic condition and $m\odot x_1$ masks the real target speech as the region to be predicted.

\textbf{3) Prompt-Token Text Conditioning:}

In pretrained F5-TTS, the text condition follows a structured layout: the prompt transcript comes before the target text, and the remaining positions are filled with filler tokens.
This layout implicitly associates text-side positions with the prompt and target regions in the acoustic sequence.
If the audio prompt transcript is directly removed during SFT, the target text is shifted into positions that previously corresponded to the audio prompt transcript, disrupting the learned text-audio correspondence and making adaptation unnecessarily difficult.

To retain the pretraining layout while moving to transcript-free prompting, we replace the missing prompt-token region with $N$ learnable prompt tokens $\langle P\rangle$ followed by an end-of-sequence (EOS) marker $\text{`. '}$ before the target text.
The learnable prompt tokens serve as explicit transcript-free prompt indicators and provide placeholders in the text sequence for the acoustic prompt.
The EOS marker reuses the text-boundary cue already learned during F5-TTS pretraining, where the sequence $\text{`. '}$ marks the end of a sentence or utterance.
Placed after the prompt tokens, this marker signals the boundary between the prompt-token region and the following target text, which helps the model adapt faster during fine-tuning.
The full text sequence $z$ is constructed as
\begin{equation}
\begin{aligned}
  z = \bigl(&\underbrace{\langle P\rangle,\ldots,\langle P\rangle}_{N},\;
             \text{`\,.\,'},\;\text{`\, \,'},\\
             &\;c_1,\ldots,c_M,\;
             \underbrace{\langle F\rangle,\ldots,\langle F\rangle}_{\tau - N - M - 2}\bigr),
\end{aligned}
\end{equation}
where $c_1,\ldots,c_M$ is the target text token sequence of length $M$ and $\langle F\rangle$ denotes filler tokens used for padding.
The number of prompt tokens $N$ is determined according to the ratio between the prompt and target speech durations.
This ratio-based design ensures that the length of the prompt-token region approximates what the prompt transcript would occupy in the pretraining text condition.

\subsection{Silence-Robust Speaking Rate Predictor}

\begin{figure*}[!t]
  \centering
  \includegraphics[width=0.75\linewidth]{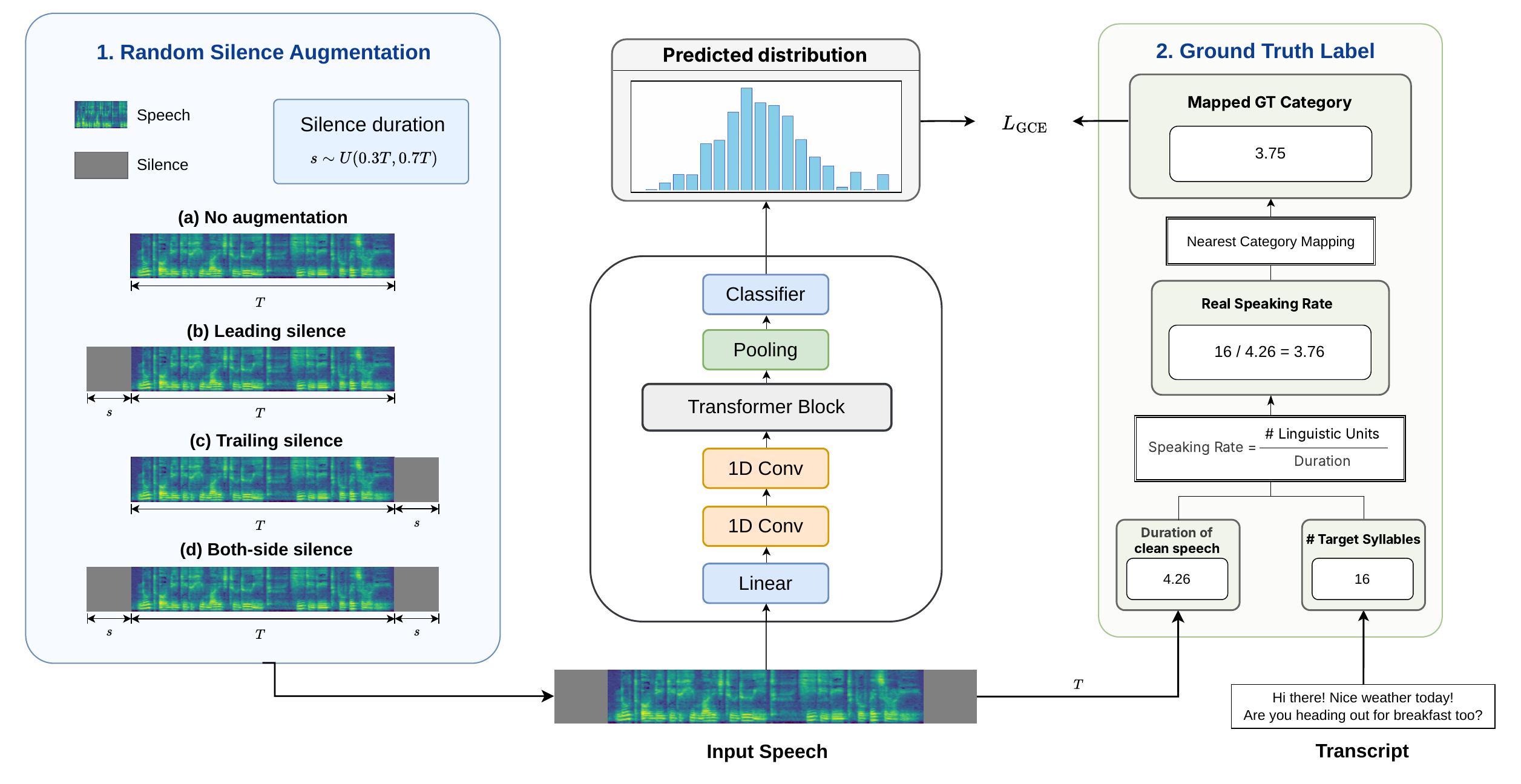}
  \caption{Silence-aware augmentation for training the silence-robust speaking rate predictor.
  With some probability, silence is inserted before, after, or on both sides of the prompt audio,
  so the predictor sees prompts with leading or trailing silence during training.
  The speaking rate label is unchanged, because the augmentation only pads the prompt with
  silence without modifying the speech segment that determines the rate.}
  \label{fig:srp_aug}
\end{figure*}

Following Cross-Lingual F5-TTS~\cite{liu2026crosslingualf5tts}, we use a syllable-level speaking rate predictor to estimate the target speech duration from the prompt audio and the target text.
However, real prompts often contain leading or trailing silence, which can bias the observed speaking rate downward and lead to overestimated target durations.
To make the predictor robust to such prompts, we apply silence-aware augmentation during training: as shown in Fig.~\ref{fig:srp_aug}, with a certain probability, silence is inserted before, after, or on both sides of the prompt audio.
The model architecture and training labels remain unchanged; only the acoustic input to the predictor is augmented.

%% file: sections/4_experiments.tex
\section{Experimental Setup}
\label{sec:experiments}

\subsection{Dataset}
\label{sec:dataset}

We use the Emilia dataset~\cite{he2024emilia} as the primary training source.
After filtering transcription failures and misclassified language speech, the training pool contains approximately 95K hours of English and Chinese speech.

\input{tables/0_training_data}

For each real utterance in Emilia, we generate a corresponding synthetic prompt with a pretrained F5-TTS model~\cite{chen2025f5tts}, yielding paired synthetic-prompt and real-target speech for fine-tuning.
To examine the effect of data scale, we additionally construct three reduced subsets by selecting utterances with the highest DNSMOS scores~\cite{reddy2022dnsmosp835} from the same training pool, balanced across the two languages, and generate synthetic prompts for each subset in the same manner.
Table~\ref{tab:training_data} summarizes the real and synthetic durations at each scale.

\subsection{Model Setup}
\label{sec:setup}

Cross-Lingual F5-TTS 2 is initialized from pretrained F5-TTS and fine-tuned on the constructed real-synthetic pairs.
The model uses the same F5-TTS-Base architecture as the pretrained model, with a diffusion transformer backbone containing 22 layers, 16 attention heads, and 1024 dimensional embeddings.
SFT is conducted for 100K updates on eight NVIDIA A100 GPUs.
We use a per-GPU batch size of 38,400 audio frames and the AdamW optimizer.
The learning rate is linearly warmed up to $7.5 \times 10^{-5}$ over the first 20K updates and then linearly decayed.

The speaking rate predictor uses a transformer-based architecture with 6 layers, 8 attention heads, and 512-dimensional embeddings.
It is trained on a balanced subset of 500 hours each from the Chinese and English portions of Emilia, following Cross-Lingual F5-TTS~\cite{liu2026crosslingualf5tts}.
Training is conducted on four A100 GPUs for 30K updates with a per-GPU batch size of 19,200 audio frames.
The learning rate is warmed up to $2.5 \times 10^{-4}$ over the first 7.5K updates and then linearly decayed.
For Gaussian Cross-Entropy loss, we set the standard deviation to $\sigma = 1.0$.
For silence data augmentation, we probabilistically insert silence into the prompt audio during training: 70\% of prompts remain unchanged, while silence is inserted at the beginning, at the end, or at both sides of the prompt with 10\% probability each.
The inserted silence duration is uniformly sampled between 30\% and 70\% of the original prompt duration.

For inference, we follow the F5-TTS setting, with the Euler ODE solver using 32 function evaluations (NFE = 32), CFG strength 2.0, sway sampling coefficient -1.0, and a pretrained Vocos vocoder~\cite{siuzdak2024vocos}.

\subsection{Baselines}
\label{sec:baselines}

We adopt F5-TTS~\cite{chen2025f5tts} and Cross-Lingual F5-TTS~\cite{liu2026crosslingualf5tts} as baselines, both introduced in Section~\ref{sec:method}.
The two baselines estimate the target duration differently at inference.
F5-TTS uses a ratio-based method, in which the duration equals the audio prompt duration scaled by the ratio of the target text length to the reference text length.
Cross-Lingual F5-TTS uses the original speaking rate predictor trained without silence augmentation.
In contrast, Cross-Lingual F5-TTS 2 estimates the target duration with our speaking rate predictor trained with silence augmentation.

\input{tables/1_baseline_results}

\subsection{Evaluation}
\label{sec:evaluation}

We follow the evaluation setting of F5-TTS, using LibriSpeech-PC \textit{test-clean}~\cite{meister2023librispeechpc} and Seed-TTS-eval~\cite{anastassiou2024seedtts} as the intra-lingual evaluation sets. 

To evaluate cross-lingual voice cloning, we further build a test set whose audio prompts come from eight languages that are out-of-distribution with respect to the English and Chinese training data: Korean from Emilia~\cite{he2024emilia}, Vietnamese from Dolly-Audio~\cite{dollyaudio}, and Thai, Indonesian, French, German, Russian, and Maltese from Common Voice~\cite{commonvoice}. These languages span different families and writing systems, covering both tonal and non-tonal as well as high-resource and less represented cases. Each prompt language is paired with 500 English and 500 Chinese target sentences drawn from Common Voice, giving 16 cross-lingual subsets in which the prompt language always differs from the target.

We evaluate TTS performance using three objective metrics:

\noindent\textbf{Word Error Rate (WER)} measures the intelligibility of synthesized speech by comparing automatic transcriptions with the ground-truth text.
We use Whisper-large-V3~\cite{radford2023whisper} for English and Paraformer-zh~\cite{gao2022paraformer} for Chinese recognition, and compute WER accordingly.

\noindent\textbf{Speaker Similarity (SIM-o)} measures speaker similarity between synthesized speech and the reference speech.
We extract speaker embeddings using a WavLM-large-based speaker verification model~\cite{chen2022wavlm} and compute their cosine similarity.

\noindent\textbf{UTMOS}~\cite{saeki2022utmos} provides an automatic estimate of speech naturalness and audio quality using a pretrained MOS prediction model.
It is used as an objective proxy for naturalness rather than a subjective listening score.

In Table~\ref{tab:baseline_results}, we further report
two subjective metrics:

\noindent\textbf{Similarity MOS (SMOS)} measures the similarity in timbre and prosody
between the synthesized speech and the speaker prompt. SMOS is on a scale of $1$ to $5$.

\noindent\textbf{Comparative MOS (CMOS)} measures the naturalness of the synthesized
speech relative to the ground-truth recording of the same utterance. CMOS is on a
scale of $-3$ to $+3$.

To assess robustness to leading and trailing silence, we additionally evaluate on a \textit{padded} prompt that adds silence to both sides of the original prompt, each side fixed to 50\% of the prompt duration, and compare it against the default \textit{clean} prompt on LibriSpeech-PC \textit{test-clean}, Seed-TTS \textit{test-en}, and Seed-TTS \textit{test-zh}. Under both prompts we report both the duration prediction accuracy of the speaking rate predictor, using the two metrics below, and the resulting TTS performance, using the WER, SIM-o, and UTMOS metrics defined above.

\noindent\textbf{Mean Relative Error (MRE)} measures the average relative difference between the predicted duration and the ground-truth duration.
The predicted duration is obtained from the number of target linguistic units divided by the predicted speaking rate.

\noindent\textbf{Mean Absolute Error (MAE)} measures the absolute deviation between the predicted and ground-truth durations in seconds.

%% file: tables/0_training_data.tex
\begin{table}[!htbp]
\renewcommand\arraystretch{1.0}
\caption{Training data for synthetic-prompt SFT at each data scale.}
\label{tab:training_data}
\centering
\footnotesize
\begin{tabular}{cc}
\toprule
\textbf{Real speech (h)} & \textbf{Synthetic speech (h)} \\
\midrule
1{,}000  & 286 \\
2{,}000  & 574 \\
5{,}000  & 1{,}443 \\
95{,}282 & 29{,}169 \\
\bottomrule
\end{tabular}
\end{table}

%% file: tables/1_baseline_results.tex
\begin{table*}[!t]
\renewcommand\arraystretch{1.0}
\caption{Objective and subjective TTS results on LibriSpeech-PC test-clean, Seed-TTS test-en, and Seed-TTS test-zh. SMOS ($1$--$5$) and CMOS (relative to ground truth) are obtained from a subjective listening test on $15$ randomly-sampled utterances per test set.}
\label{tab:baseline_results}
\centering
\footnotesize
\begin{tabular}{lccccc}
\toprule
\textbf{System} & \textbf{WER(\%)$\downarrow$} & \textbf{SIM-o$\uparrow$} & \textbf{UTMOS$\uparrow$} & \textbf{SMOS$\uparrow$} & \textbf{CMOS$\uparrow$} \\
\midrule
\multicolumn{6}{c}{\textbf{LibriSpeech-PC \textit{test-clean}}} \\
\midrule
F5-TTS & 2.205 & 0.668 & 3.797 & \textbf{4.324} & \textbf{+0.480} \\
Cross-Lingual F5-TTS & 2.144 & 0.658 & \textbf{3.893} & 4.063 & +0.258 \\
\textbf{Cross-Lingual F5-TTS 2} & \textbf{2.014} & \textbf{0.687} & 3.704 & 4.216 & +0.150 \\
\midrule
\multicolumn{6}{c}{\textbf{Seed-TTS \textit{test-en}}} \\
\midrule
F5-TTS & \textbf{1.545} & 0.676 & 3.582 & \textbf{4.148} & \textbf{+0.141} \\
Cross-Lingual F5-TTS & 1.613 & 0.659 & \textbf{3.635} & 4.051 & +0.003 \\
\textbf{Cross-Lingual F5-TTS 2} & 1.629 & \textbf{0.683} & 3.544 & 4.082 & -0.064 \\
\midrule
\multicolumn{6}{c}{\textbf{Seed-TTS \textit{test-zh}}} \\
\midrule
F5-TTS & \textbf{1.475} & 0.762 & 2.898 & \textbf{4.290} & +0.508 \\
Cross-Lingual F5-TTS & 1.521 & 0.763 & \textbf{2.901} & 4.064 & \textbf{+0.677} \\
\textbf{Cross-Lingual F5-TTS 2} & 1.594 & \textbf{0.768} & 2.761 & 4.098 & -0.068 \\
\bottomrule
\end{tabular}
\end{table*}

%% file: sections/5_results.tex
\section{Results and Analysis}
\label{sec:results}

\subsection{Main Results}
\label{sec:baseline_results}

Table~\ref{tab:baseline_results} compares Cross-Lingual F5-TTS~2 with F5-TTS and Cross-Lingual F5-TTS on LibriSpeech-PC \textit{test-clean} and Seed-TTS test sets along three dimensions: intelligibility, similarity, and naturalness. Our method achieves competitive performance across all three test sets. In intelligibility, Cross-Lingual F5-TTS~2 is comparable to both baselines on WER and obtains the lowest value on LibriSpeech-PC \textit{test-clean} ($2.014\%$), indicating that replacing the real audio prompt does not degrade intelligibility. In similarity, Cross-Lingual F5-TTS~2 achieves the highest SIM-o on all three test sets, with the largest gains on the English benchmarks (LibriSpeech-PC \textit{test-clean} and Seed-TTS \textit{test-en}) and a slight improvement on Seed-TTS \textit{test-zh}. This gain comes from the synthetic prompt used during supervised fine-tuning, which provides a cleaner conditioning signal: real recordings carry unstable prosody and paralinguistic information that the model must discount before copying the target voice, whereas the synthetic prompt is steadier and plainer, so the model learns speaker identity more reliably. Cross-Lingual F5-TTS~2 also obtains a higher SMOS than Cross-Lingual F5-TTS but a slightly lower one than F5-TTS on all three test sets. Because SMOS reflects both timbre and prosody while SIM-o captures mainly timbre, the flatter prosody induced by the synthetic prompt lowers SMOS relative to F5-TTS. In naturalness, Cross-Lingual F5-TTS~2 is slightly lower than both baselines on UTMOS and CMOS on all three test sets, as the cleaner but prosodically flatter synthetic prompt biases the model. This modest decrease in naturalness is an acceptable trade-off for the substantial gain in speaker similarity. Overall, Cross-Lingual F5-TTS~2 removes the dependency on audio prompt transcripts while improving speaker similarity and maintaining intelligibility, at only a modest cost in naturalness.

\subsection{Cross-Lingual Voice Cloning Results}
\label{sec:crosslingual}

\input{tables/4_crosslingual}

Table~\ref{tab:crosslingual} evaluates cross-lingual voice cloning, where the audio prompt is in one of eight out-of-distribution languages and the target text is English or Chinese, so the prompt language always differs from the target. We compare Cross-Lingual F5-TTS 2 with Cross-Lingual F5-TTS across all 16 prompt-target pairs.

The overall pattern is consistent with the intra-lingual results in Section~\ref{sec:baseline_results}: Cross-Lingual F5-TTS 2 obtains higher SIM-o on 14 of the 16 pairs; Cross-Lingual F5-TTS obtains higher UTMOS, and WER is comparable between the two models.

These results show that Cross-Lingual F5-TTS 2 can take an out-of-distribution audio prompt, without access to its transcript, and clone the speaker identity into English or Chinese across a wide range of languages, including less represented ones such as Maltese, whose Semitic origins place it outside the Indo-European and Sino-Tibetan families of the training languages. This is the practical benefit of removing the transcript dependency: the model can clone speakers from prompt languages for which a reliable transcript is difficult to obtain.

\input{tables/2_text_conditioning_ablation}

\subsection{Ablation Study on Text Conditioning}
\label{sec:text_conditioning_ablation}

\begin{figure}[!h]
\centering
\includegraphics[width=\linewidth]{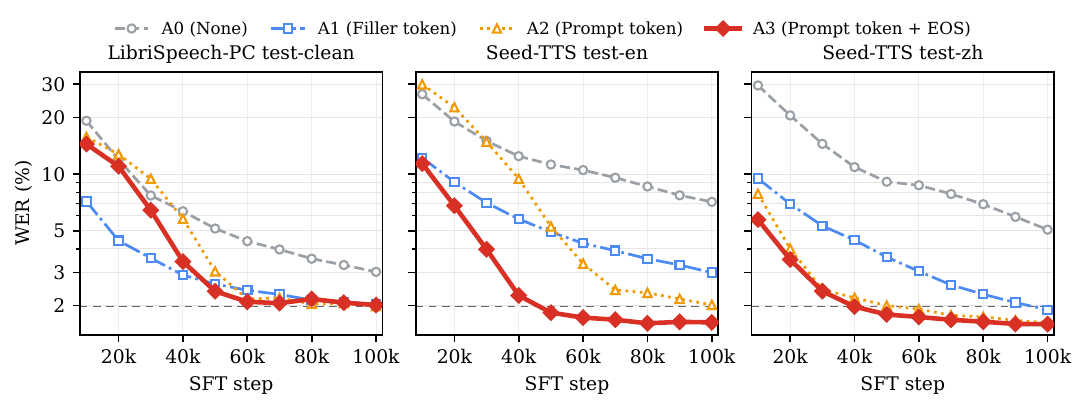}
\caption{WER versus SFT step for the four text-conditioning variants in Table~\ref{tab:text_conditioning_ablation}, on LibriSpeech-PC \textit{test-clean} (left), Seed-TTS \textit{test-en} (middle), and \textit{test-zh} (right); the y-axis is log-scaled and the dashed line marks a 2\% WER reference.}
\label{fig:wer_convergence}
\end{figure}

Table~\ref{tab:text_conditioning_ablation} examines how the text conditioning that replaces the missing audio prompt transcript affects fine-tuning. We compare four variants, each fine-tuned for 100K steps: no placeholder (A0), $N$ filler tokens (A1), $N$ learnable prompt tokens (A2), and $N$ learnable prompt tokens followed by an EOS marker (A3), where $N$ is set from the prompt-to-target duration ratio as described in Section~\ref{sec:method}. The placeholder column denotes the token used in place of the audio prompt transcript, and the EOS column denotes whether an EOS marker is appended. Fig.~\ref{fig:wer_convergence} plots WER against the number of SFT steps on the three test sets, with a log-scaled y-axis and a dashed line at 2\% WER as a reference, which makes the convergence speed of each variant visible.

Among the four variants, A3 converges fastest on the Seed-TTS sets, where it crosses the 2\% reference earliest and reaches the lowest final WER. The comparison among A0, A1, and A2 isolates the effect of the placeholder. Both A1 and A2 fill the dropped prompt-token region with $N$ tokens and converge to a much lower WER than removing it outright (A0), because they preserve the text-conditioning layout seen during pretraining; leaving the region empty (A0) shifts the target text into positions that previously held the audio prompt transcript, abruptly changing the structure and perturbing the attention distribution that the pretrained model relies on. By comparing A2 with A1, A2 outperforms A1 because they form an explicit, trainable signal: each prompt token is updated by gradient descent into a task-specific representation. In contrast, filler tokens are masked out during the forward pass and provide no such signal, leaving the prompt-token region without a text-side anchor. A2 also exhibits a distinctive slow-start phase, during which its randomly initialized prompt tokens provide no useful conditioning until training shapes them and WER improves only slowly before falling sharply. This phase lasts longest on Seed-TTS \textit{test-en} and is shortest on Seed-TTS \textit{test-zh}, where A2 is already among the lowest at 10K steps.

Comparing A3 with A2 isolates the effect of the EOS marker. As shown in Fig.~\ref{fig:wer_convergence}, appending the EOS marker mainly accelerates convergence: A3 converges far faster than A2 on Seed-TTS \textit{test-en} and settles at a lower final WER, while it converges slightly faster and reaches a comparable final WER on LibriSpeech-PC \textit{test-clean} and Seed-TTS \textit{test-zh}. The EOS marker accelerates convergence because F5-TTS has already learned during pretraining that the sequence `.\,' marks the end of an utterance, so reusing it at fine-tuning time supplies a familiar cue that closes the prompt-token region and signals where the target segment begins.

\input{tables/3_data_efficiency}

\subsection{Ablation Study on Fine-Tuning Data Scale}
\label{sec:data_scale}

Table~\ref{tab:data_scale} studies how the amount of fine-tuning data affects synthetic-prompt SFT. We fine-tune Cross-Lingual F5-TTS 2 at four data scales, pairing each real utterance with a synthetic prompt as in Table~\ref{tab:training_data}. Overall, more data brings only a small improvement, and even the smallest scale (B0) already reaches performance close to the full set.

The three metrics respond to scale differently. SIM-o is the only one that improves consistently with more data: B3 obtains the highest score on all three test sets, although the margin over the smaller scales is small (within 0.005). WER and UTMOS show no clear trend and vary within a narrow band; B3 is not uniformly best, and B1 even reaches the lowest WER on Seed-TTS \textit{test-zh}.

This limited sensitivity to data scale is expected. Cross-Lingual F5-TTS 2 is initialized from a pretrained F5-TTS that has already learned acoustic modeling, so fine-tuning only needs to adapt the model to the transcript-free text conditioning rather than to relearn the acoustic mapping. A moderate amount of data is sufficient for this adaptation: a few thousand hours recover most of the benefit, and the full pool adds only a marginal gain in speaker similarity.

\input{tables/5_srp}

\subsection{Robustness to Prompt Silence}
\label{sec:srp_robustness}

Tables~\ref{tab:srp_duration} and~\ref{tab:srp_robust} examine how robust the duration estimation is to silence in the audio prompt, under a clean prompt and a padded prompt that adds leading and trailing silence. Both tables compare the original speaking rate predictor from Cross-Lingual F5-TTS, with the silence-robust speaking rate predictor of Section~\ref{sec:method}. Table~\ref{tab:srp_duration} reports the duration accuracy of each predictor, and Table~\ref{tab:srp_robust} reports the TTS performance when its predicted durations are used by the same Cross-Lingual F5-TTS 2 model, so that the two settings differ only in the estimated target duration.

\input{tables/6_srp_robust}

Under clean prompts the two predictors behave almost identically: silence augmentation leaves MAE and MRE essentially unchanged, with MRE rising by only 0.6 and 0.8 points on Seed-TTS \textit{test-en} and \textit{test-zh}, so it costs little when the prompt is already clean. The difference emerges under padded prompts. Once silence is added in the audio prompt, the original predictor fails badly, with MRE above 70\% and MAE exceeding three seconds, because the silence lowers the predicted speaking rate and the estimated duration is inflated accordingly. The silence-robust speaking rate predictor instead stays close to its clean-prompt accuracy.

Table~\ref{tab:srp_robust} shows how these duration errors carry over to the synthesized speech. On clean prompts the durations from the two predictors lead to comparable WER, SIM-o, and UTMOS. On padded prompts the durations from the original predictor degrade all three metrics, with higher WER, lower SIM-o, and lower UTMOS. The cause is the inflated duration: when the predicted speaking rate is too slow, the target is allotted more frames than the content needs, the speech is stretched, and its prosody is distorted, which lowers both intelligibility and naturalness. With durations from the silence-robust speaking rate predictor, the padded-prompt results stay close to the clean-prompt ones, in line with the accurate durations in Table~\ref{tab:srp_duration}.

%% file: tables/4_crosslingual.tex
\begin{table*}[!t]
\renewcommand\arraystretch{1.0}
\caption{Cross-lingual results with out-of-distribution (OOD) audio prompts. The prompt language (ko/th/vi/id/fr/de/ru/mt) differs from the target language (en/zh) and is unseen during training. WER (\%), SIM-o, and UTMOS are reported for each language pair.}
\label{tab:crosslingual}
\centering
\footnotesize
\begin{tabular}{l *{6}{>{\centering\arraybackslash}p{0.95cm}}}
\toprule
\multirow{2}{*}{\textbf{Prompt Language}} & \multicolumn{3}{c}{\textbf{Cross-Lingual F5-TTS}} & \multicolumn{3}{c}{\textbf{Cross-Lingual F5-TTS 2}} \\
\cmidrule(lr){2-4} \cmidrule(lr){5-7}
 & \textbf{WER$\downarrow$} & \textbf{SIM-o$\uparrow$} & \textbf{UTMOS$\uparrow$} & \textbf{WER$\downarrow$} & \textbf{SIM-o$\uparrow$} & \textbf{UTMOS$\uparrow$} \\
\midrule
\multicolumn{7}{c}{\textbf{Target Language: English}} \\
\midrule
Korean & 3.496 & 0.449 & \textbf{3.426} & \textbf{3.383} & \textbf{0.455} & 3.370 \\
Thai & \textbf{3.599} & 0.468 & \textbf{3.705} & 4.236 & \textbf{0.508} & 3.481 \\
Vietnamese & \textbf{3.281} & \textbf{0.510} & \textbf{4.184} & 3.464 & 0.503 & 4.153 \\
Indonesian & \textbf{4.330} & 0.431 & \textbf{3.405} & 4.442 & \textbf{0.450} & 3.290 \\
French & 3.929 & 0.506 & \textbf{3.718} & \textbf{3.841} & \textbf{0.532} & 3.599 \\
German & 3.913 & 0.554 & \textbf{3.602} & \textbf{3.431} & \textbf{0.579} & 3.509 \\
Russian & 4.337 & 0.492 & \textbf{3.608} & \textbf{3.614} & \textbf{0.517} & 3.513 \\
Maltese & 3.884 & 0.516 & \textbf{3.693} & \textbf{3.486} & \textbf{0.531} & 3.622 \\
\midrule
\multicolumn{7}{c}{\textbf{Target Language: Chinese}} \\
\midrule
Korean & \textbf{4.133} & 0.644 & \textbf{2.761} & 4.574 & \textbf{0.658} & 2.602 \\
Thai & \textbf{3.938} & 0.607 & \textbf{2.888} & 5.030 & \textbf{0.626} & 2.609 \\
Vietnamese & \textbf{4.268} & \textbf{0.673} & \textbf{3.550} & 4.517 & 0.671 & 3.488 \\
Indonesian & 5.370 & 0.478 & \textbf{2.788} & \textbf{4.714} & \textbf{0.500} & 2.593 \\
French & 4.961 & 0.515 & \textbf{3.097} & \textbf{4.933} & \textbf{0.529} & 2.862 \\
German & \textbf{4.685} & 0.512 & \textbf{2.950} & 4.771 & \textbf{0.532} & 2.777 \\
Russian & 5.325 & 0.544 & \textbf{2.976} & \textbf{4.659} & \textbf{0.553} & 2.814 \\
Maltese & \textbf{4.842} & 0.518 & \textbf{3.002} & 5.332 & \textbf{0.529} & 2.845 \\
\bottomrule
\end{tabular}
\end{table*}

%% file: tables/2_text_conditioning_ablation.tex
\begin{table}[!htbp]
\renewcommand\arraystretch{1.0}
\caption{Ablation of prompt-token text conditioning. The placeholder denotes the substitute for the missing prompt transcript.}
\label{tab:text_conditioning_ablation}
\centering
\footnotesize
\begin{tabular}{llcccc}
\toprule
\textbf{ID} & \textbf{Placeholder} & \textbf{EOS} & \textbf{WER(\%)$\downarrow$} & \textbf{SIM-o$\uparrow$} & \textbf{UTMOS$\uparrow$} \\
\midrule
\multicolumn{6}{c}{\textbf{LibriSpeech-PC \textit{test-clean}}} \\
\midrule
A0 & None & \xmark & 3.022 & 0.686 & 3.705 \\
A1 & Filler token & \xmark & 2.028 & 0.687 & \textbf{3.707} \\
A2 & Prompt token & \xmark & \textbf{1.984} & \textbf{0.688} & 3.703 \\
A3 & Prompt token & \cmark & 2.014 & 0.687 & 3.704 \\
\midrule
\multicolumn{6}{c}{\textbf{Seed-TTS \textit{test-en}}} \\
\midrule
A0 & None & \xmark & 7.116 & 0.675 & 3.516 \\
A1 & Filler token & \xmark & 2.994 & 0.681 & 3.514 \\
A2 & Prompt token & \xmark & 2.018 & 0.683 & 3.508 \\
A3 & Prompt token & \cmark & \textbf{1.629} & \textbf{0.683} & \textbf{3.544} \\
\midrule
\multicolumn{6}{c}{\textbf{Seed-TTS \textit{test-zh}}} \\
\midrule
A0 & None & \xmark & 5.066 & 0.765 & 2.745 \\
A1 & Filler token & \xmark & 1.900 & 0.767 & \textbf{2.777} \\
A2 & Prompt token & \xmark & 1.621 & \textbf{0.768} & 2.760 \\
A3 & Prompt token & \cmark & \textbf{1.594} & 0.768 & 2.761 \\
\bottomrule
\end{tabular}
\end{table}

%% file: tables/3_data_efficiency.tex
\begin{table}[!htbp]
\renewcommand\arraystretch{1.0}
\caption{Ablation study on fine-tuning data scale for synthetic-prompt SFT.}
\label{tab:data_scale}
\centering
\footnotesize
\begin{tabular}{llccc}
\toprule
\textbf{ID} & \textbf{Real / Synthetic (h)} & \textbf{WER(\%)$\downarrow$} & \textbf{SIM-o$\uparrow$} & \textbf{UTMOS$\uparrow$} \\
\midrule
\multicolumn{5}{c}{\textbf{LibriSpeech-PC \textit{test-clean}}} \\
\midrule
B0 & 1{,}000 / 286 & 2.223 & 0.683 & 3.747 \\
B1 & 2{,}000 / 574 & 2.264 & 0.683 & 3.739 \\
B2 & 5{,}000 / 1{,}443 & 2.190 & 0.686 & \textbf{3.748} \\
B3 & 95{,}282 / 29{,}169 & \textbf{2.014} & \textbf{0.687} & 3.704 \\
\midrule
\multicolumn{5}{c}{\textbf{Seed-TTS \textit{test-en}}} \\
\midrule
B0 & 1{,}000 / 286 & 1.760 & 0.679 & 3.533 \\
B1 & 2{,}000 / 574 & 1.645 & 0.679 & 3.526 \\
B2 & 5{,}000 / 1{,}443 & 1.690 & 0.680 & 3.538 \\
B3 & 95{,}282 / 29{,}169 & \textbf{1.629} & \textbf{0.683} & \textbf{3.544} \\
\midrule
\multicolumn{5}{c}{\textbf{Seed-TTS \textit{test-zh}}} \\
\midrule
B0 & 1{,}000 / 286 & 1.591 & 0.765 & \textbf{2.770} \\
B1 & 2{,}000 / 574 & \textbf{1.552} & 0.763 & 2.759 \\
B2 & 5{,}000 / 1{,}443 & 1.589 & 0.765 & 2.769 \\
B3 & 95{,}282 / 29{,}169 & 1.594 & \textbf{0.768} & 2.761 \\
\bottomrule
\end{tabular}
\end{table}

%% file: tables/5_srp.tex
\begin{table}[!htbp]
\renewcommand\arraystretch{1.0}
\caption{Duration prediction accuracy of the original and silence-robust speaking rate predictors.}
\label{tab:srp_duration}
\centering
\footnotesize
\begin{tabular}{l cc cc}
\toprule
\multirow{2}{*}{\textbf{Duration Method}} & \multicolumn{2}{c}{\textbf{Clean Prompt}} & \multicolumn{2}{c}{\textbf{Padded Prompt}} \\
\cmidrule(lr){2-3} \cmidrule(lr){4-5}
 & \textbf{MAE(s)$\downarrow$} & \textbf{MRE(\%)$\downarrow$} & \textbf{MAE(s)$\downarrow$} & \textbf{MRE(\%)$\downarrow$} \\
\midrule
\multicolumn{5}{c}{\textbf{LibriSpeech-PC \textit{test-clean}}} \\
\midrule
Syllable-level SRP & \textbf{0.729} & \textbf{11.414} & 4.495 & 70.211 \\
\quad\textbf{+ Silence Aug.} & \textbf{0.729} & 11.567 & \textbf{0.817} & \textbf{13.154} \\
\midrule
\multicolumn{5}{c}{\textbf{Seed-TTS \textit{test-en}}} \\
\midrule
Syllable-level SRP & \textbf{0.683} & \textbf{15.774} & 3.034 & 70.984 \\
\quad\textbf{+ Silence Aug.} & 0.697 & 16.330 & \textbf{0.652} & \textbf{15.364} \\
\midrule
\multicolumn{5}{c}{\textbf{Seed-TTS \textit{test-zh}}} \\
\midrule
Syllable-level SRP & \textbf{0.730} & \textbf{12.692} & 4.708 & 85.491 \\
\quad\textbf{+ Silence Aug.} & 0.772 & 13.532 & \textbf{1.034} & \textbf{18.933} \\
\bottomrule
\end{tabular}
\end{table}

%% file: tables/6_srp_robust.tex
\begin{table*}[!htbp]
\renewcommand\arraystretch{1.0}
\caption{TTS performance of Cross-Lingual F5-TTS 2 using target durations from the original and silence-robust speaking rate predictors under Clean and Padded prompts.}
\label{tab:srp_robust}
\centering
\footnotesize
\begin{tabular}{l ccc ccc}
\toprule
\multirow{2}{*}{\textbf{Duration Method}} & \multicolumn{3}{c}{\textbf{Clean Prompt}} & \multicolumn{3}{c}{\textbf{Padded Prompt}} \\
\cmidrule(lr){2-4} \cmidrule(lr){5-7}
 & \textbf{WER(\%)$\downarrow$} & \textbf{SIM-o$\uparrow$} & \textbf{UTMOS$\uparrow$} & \textbf{WER(\%)$\downarrow$} & \textbf{SIM-o$\uparrow$} & \textbf{UTMOS$\uparrow$} \\
\midrule
\multicolumn{7}{c}{\textbf{LibriSpeech-PC \textit{test-clean}}} \\
\midrule
Syllable-level SRP & \textbf{1.979} & 0.683 & \textbf{3.704} & 3.463 & 0.672 & 3.656 \\
\quad\textbf{+ Silence Aug.} & 2.014 & \textbf{0.687} & \textbf{3.704} & \textbf{2.454} & \textbf{0.685} & \textbf{3.731} \\
\midrule
\multicolumn{7}{c}{\textbf{Seed-TTS \textit{test-en}}} \\
\midrule
Syllable-level SRP & \textbf{1.626} & 0.675 & \textbf{3.550} & 3.241 & 0.674 & 3.477 \\
\quad\textbf{+ Silence Aug.} & 1.629 & \textbf{0.683} & 3.544 & \textbf{1.919} & \textbf{0.675} & \textbf{3.559} \\
\midrule
\multicolumn{7}{c}{\textbf{Seed-TTS \textit{test-zh}}} \\
\midrule
Syllable-level SRP & \textbf{1.494} & 0.766 & \textbf{2.802} & 4.006 & 0.725 & 2.746 \\
\quad\textbf{+ Silence Aug.} & 1.594 & \textbf{0.768} & 2.761 & \textbf{1.738} & \textbf{0.753} & \textbf{2.870} \\
\bottomrule
\end{tabular}
\end{table*}

%% file: sections/6_conclusion.tex
\section{Conclusion}
\label{sec:conclusion}
In this work, we present Cross-Lingual F5-TTS 2, a simplified fine-tuning framework that achieves cross-lingual zero-shot voice cloning without audio prompt transcripts or forced alignment. Our method pairs each real utterance with a same-speaker prompt synthesized by a pretrained F5-TTS, and fine-tunes the same pretrained model on these pairs. For duration estimation, we follow Cross-Lingual F5-TTS and use a syllable-level speaking rate predictor, which we make robust to leading or trailing silence through silence-aware augmentation. Experiments show that Cross-Lingual F5-TTS 2 reaches higher speaker similarity than F5-TTS and Cross-Lingual F5-TTS while maintaining intelligibility, and generalizes to out-of-distribution prompt languages without their transcripts. The silence-robust speaking rate predictor further keeps the predicted duration accurate when the prompt contains leading or trailing silence. The main limitation is a small drop in naturalness, caused by the prosodically flatter synthetic prompt. In future work, we will improve the expressiveness of the synthetic prompt and extend the approach to more languages beyond English and Chinese.